\documentclass[pdflatex,aps,prl,nofootinbib,twocolumn,showpacs,superscriptaddress]{revtex4-1}
\usepackage{amssymb,amsmath,color,hyperref,graphicx,multirow,tabularx,physics}
\usepackage{davitex}

\newcommand{\Pbare}{P^{\text{bare}}}

\newcommand{\freg}{f_{\rm reg}}
\newcommand{\Hmax}{H_{\rm max}}

\begin{document}


\title{Sensitivity of the Polyakov loop and related observables to chiral symmetry restoration} 


\author{D. A. Clarke}
\affiliation{Fakult\"at f\"ur Physik, Universit\"at Bielefeld, D-33615
Bielefeld, Germany}
\author{O. Kaczmarek}
\affiliation{ Key Laboratory of Quark \& Lepton Physics (MOE) and Institute
of Particle Physics, Central China Normal University, Wuhan 430079, China}
\affiliation{Fakult\"at f\"ur Physik, Universit\"at Bielefeld, D-33615
Bielefeld, Germany}
\author{F. Karsch}
\affiliation{Fakult\"at f\"ur Physik, Universit\"at Bielefeld, D-33615
Bielefeld, Germany}
\author{Anirban Lahiri}
\affiliation{Fakult\"at f\"ur Physik, Universit\"at Bielefeld, D-33615
Bielefeld, Germany}
\author{Mugdha Sarkar}
\affiliation{Fakult\"at f\"ur Physik, Universit\"at Bielefeld, D-33615
Bielefeld, Germany}

\date{\today}

\begin{abstract}

While the Polyakov loop is an order parameter of the deconfinement transition
in the heavy quark mass regime of QCD, its sensitivity to the deconfinement of
light, dynamical quarks in QCD is not apparent. On the other hand, the 
quark mass dependence of the Polyakov loop is sensitive to the appearance
of a chiral phase transition. 
Using lattice QCD calculations in the staggered fermion discretization 
scheme at finite values of the lattice spacing, $aT=1/8$, we
show here, for the first time, that the Polyakov loop expectation
value, and the heavy quark free energy extracted from it, behave like 
energy-like observables in the vicinity of the chiral phase
transition temperature $T_c$. Consistent with scaling behavior of 
energy-like observables in the 3-$d$, $\O(2)$ universality class,
the quark mass derivatives  
diverge in the chiral limit at $T_c$ while the temperature derivatives
stay finite. The latter will develop a characteristic 
spike at $T_c$. This, however, may be resolved only in calculations with quark masses being two orders of magnitude
smaller than those currently accessible in lattice QCD calculations. 

\end{abstract}

\pacs{11.10.Wx, 11.15.Ha, 12.38.Aw, 12.38.Gc, 12.38.Mh, 24.60.Ky, 25.75.Gz, 25.75.Nq}

\maketitle

\emph{Introduction.---}
The Lagrangian of quantum chromodynamics (QCD), the theory describing
interactions controlled by the strong force, possesses 
exact global symmetries only in the massless (chiral) and infinite quark mass 
(pure gauge) limits. 
The latter case has been extensively exploited 
in lattice QCD calculations 
to discuss the deconfinement
phase transition in pure gauge theories 
\cite{Yaffe:1982qf,McLerran:1980pk,Kuti:1980gh} and its imprint 
in the heavy quark
sector of QCD. Also in the light quark mass region the rapid change 
of the Polyakov loop expectation value, $\ev{P}$, as function of the 
temperature, characterized by an inflection point in its $T$-dependence,
is often taken as an indication for the occurrence of deconfinement.
However, studies with (almost) physical light up and down quark masses and
improved discretization schemes for the QCD Lagrangian,
performed closer to the continuum limit,
in general show that the QCD transition is a smooth crossover, and no evidence
for an inflection point in the vicinity of the chiral transition
temperature is found
\cite{Aoki:2009sc,Borsanyi:2010bp,Bazavov:2016uvm,Clarke:2019tzf}.

In the limit of  vanishing values of the two light quark masses 
the chiral flavor symmetry, $\SU(2)_L\times \SU(2)_R$, gets restored above a 
temperature $T_c$, giving rise to a chiral phase transition 
\cite{Pisarski:1983ms}.
In QCD the non-zero light quark 
masses, $m_l$, are small on the scale of relevant temperatures, 
{\it e.g.} $T_c$. 
Thus the light quark chiral condensate, $\ev{\bar{\psi}\psi}$, is a good 
indicator for the occurrence of a phase transition in the chiral limit of QCD. 
The maxima in either the quark mass or temperature derivatives of
$\ev{\bar{\psi}\psi}$ diverge in the chiral limit, and for 
$m_l>0$ the positions of these maxima define pseudo-critical 
temperatures that converge to $T_c$ in the chiral limit. 

The Polyakov loop is a purely gluonic observable that is 
trivially invariant under chiral transformations in the fermion sector of the 
QCD Lagrangian. As far as critical behavior close to a second 
order phase transition point is concerned, it thus may be expected that 
$\ev{P}$ as well as the heavy quark free energy, $F_q=-T\ln \ev{P}$,
behave like any other ``energy-like" operator that may appear in an
effective Hamiltonian describing {\it e.g.} QCD thermodynamics in the
vicinity of the chiral phase transition. One may thus expect that 
$\ev{P}$ as well as $F_q/T$ are sensitive to critical 
behavior arising from this transition. 

Here we present results on the temperature and quark mass dependence of 
$\ev{P}$ and $F_q/T$ close to the chiral limit. We show that both of them
reflect properties of ``energy-like" observables in the vicinity of $T_c$
and discuss the resulting chiral limit behavior. 
As we will perform calculations at a non-zero value of the lattice 
spacing, using the Highly Improved Staggered Quark (HISQ) action, the relevant 
symmetry group for the discussion of universal scaling properties is the 
$\O(2)$ rather than $\O(4)$ group as it will be the case in the 
continuum limit of 
lattice QCD. However, as will become clear in the following, none of the
qualitative features that will arise from the energy-like behavior of the
Polyakov loop in theories with global $\O(N)$ symmetry will depend on this
difference.


\emph{Polyakov loop and the heavy quark free energy.---}
For lattice QCD in a finite Euclidean space-time volume 
$N_\sigma^3\times N_\tau$, the Polyakov loop, $P_{\vec{x}}$, and its
spatial average, $P$, 
\begin{equation}
\Pbare_{\vec{x}}\equiv\frac{1}{3}\tr \prod_\tau U_4\left(\vec{x},\tau\right) 
\; ,\;
  P\equiv\frac{1}{N_\sigma^3} {\rm e}^{N_\tau c(g^2)}  
  \sum_{\vec{x}}\Pbare_{\vec{x}},\nonumber
\end{equation}
are given in terms of $\SU(3)$-valued field variables, 
$U_4\left(\vec{x},\tau\right)$, 
defined on the temporal link ($\mu= 4$) originating at a Euclidean 
space-time point $(\vec{x},\tau)$. The volume $V=(N_\sigma a)^3$ and inverse 
temperature  $T^{-1}=N_\tau a$ are given in terms of the lattice spacing $a$.
The bare Polyakov loop, $\Pbare$, has been renormalized
using renormalization constants, $c(g^2)$,
determined in Ref.~\cite{Bazavov:2016uvm} (Table V) for the regularization
scheme used also in this work.

The heavy quark free energy, $F_q/T$, 
characterizes the behavior of correlation functions between static quark 
and anti-quark sources at infinite distances \cite{McLerran:1981pb},
\begin{eqnarray}\label{eq:Fav}
F_q(T,H)  &=& -T\ln \ev{P} 
= -\frac{T}{2} \lim_{\left|\vec{x}-\vec{y}\tinysp\right|\rightarrow \infty}
\ln \langle P^{\phantom\dagger}_{\vec{x}} P^\dagger_{\vec{y}} \rangle 
\; . 
\end{eqnarray}
Here $H=m_l/m_s$ parametrizes the quark mass dependence in terms
of the ratio of degenerate light quark masses $m_l\equiv m_u=m_d$ and 
the strange quark mass $m_s$.
For the analysis of the quark mass dependence of the heavy quark free
energy, we 
calculate the re\-normalization-scheme-independent mixed susceptibility
\begin{eqnarray}
\frac{\partial F_q(T,H)/T}{\partial H} &=& -
\frac{1}{\ev{P}} \frac{\partial \ev{P}}{\partial H} \equiv
- \frac{\chi_{mP}}{\ev{P}} \; ,
\label{FqH}
\end{eqnarray}
with the quark mass derivative of $\ev{P}$ given by
\begin{eqnarray}\label{eq:Ptimespbp}
\chi_{mP} &\equiv& 
\frac{\partial \ev{P}}{\partial H}
=   \ev{P  \cdot \Psi} - \ev{P} \ev{\Psi}  \; .
\end{eqnarray}
Here $\Psi\equiv\frac{1}{2} \hat{m}_s \tr M_l^{-1}$ denotes the extensive 
observable
defining a dimensionless combination of the 2-flavor light quark chiral 
condensate in terms of the light quark, staggered fermion matrix $M_l$;  
$\hat{m}_s$ is the bare strange quark mass in lattice units, 
$m_s/T= \hat{m}_s N_\tau$.

Using fit results for $F_q/T$ and its derivative with respect to the 
quark mass we will also be able to determine the derivatives of 
$\ev{P}$ and $F_q/T$ with respect to $T$. 
Similar to the $H$-derivatives, these derivatives are closely related to each other,
\begin{eqnarray}
T_c \frac{\partial F_q(T,H)/T}{\partial T} &=& -\
\frac{T_c}{\ev{P}} \frac{\partial \ev{P} }{\partial T} \; .
\label{FqT}
\end{eqnarray}

\noindent
\emph{Polyakov loop and chiral symmetry restoration.---}
Within Wilson's renormalization group approach 
\cite{Wilson:1971bg,Wilson:1971dh}, thermodynamics in the vicinity of a critical
point can be described by an effective Hamiltonian, which is defined in a
multi-dimensional space of operators (observables). These operators may be
invariant under the global symmetry that gets broken at the critical point or 
may break this symmetry explicitly. In the former case the operator is said to 
be energy-like, while in the latter case it is magnetization-like. 
In QCD the 2-flavor, light quark chiral
condensate is a typical magnetization-like operator.

The Polyakov loop is invariant 
under chiral transformations of the quark fields. Its expectation value,
$\ev{P}$, as well as the heavy quark free energy, $F_q/T$,
thus are energy-like observables.
We expect that they are sensitive to the chiral phase transition to
the extent that they receive non-analytic (singular) contributions in 
addition to analytic (regular) 
terms. 
In the vicinity of the critical point, {\it i.e.} close to $(T,H)=(T_c,0)$,
non-analytic contributions are universal scaling functions of a scaling variable
$z=z_0 t  H^{-1/\beta\delta}$ with $t=(T-T_c)/T_c$ and $z_0,\ T_c$ being 
non-universal constants. Energy-like observables receive contributions from the
scaling function, 
$f'_f(z)={\rm d}f_f(z)/{\rm d}z$, of the 3-$d$, $\O(N)$ universality class 
\cite{Engels:2011km}, which is the derivative of the scaling function $f_f(z)$ that characterizes the singular part of the logarithm of the partition function. 
For the heavy quark free energy we use the scaling ansatz,
\begin{equation}
		F_q(T,H)/T = A H^{(1-\alpha)/\beta\delta} f'_f(z) 
		+f_{\rm reg}(T,H) \; ,
\label{Fqcritical}
\end{equation}
where $A$ is another non-universal constant and
the critical exponents are
$\beta$, $\delta$ and $\alpha=2-\beta (1+\delta)$.
The regular contribution is an analytic function, which close 
to $(T_c,0)$ can be given as a Taylor series with even powers 
in $H$,
\begin{equation}
                \freg(T,H)= \sum_{i,j}  a^r_{i,2j}\ t^i H^{2j} 
                \equiv \sum_j p^r_{2j}(T) H^{2j}\; .
\label{freg}
\end{equation}

Using eqs.~\eqref{eq:Fav} and \eqref{Fqcritical} the Polyakov loop expectation 
value may be written as
\begin{equation}
		\ev{P}_{T,H} = \exp\left(-
		A H^{(1-\alpha)/\beta\delta} f'_f(z) 
		-f_{\rm reg}(T,H) \right) \; .
		\label{Pcritical}
\end{equation}
Note that for  $(T,H)$ close to $(T_c,0)$
eq.~\eqref{Pcritical} reduces to the usual 
non-exponential scaling ansatz.

\begin{figure*}[t]
\centering
\includegraphics[width=0.32\textwidth]{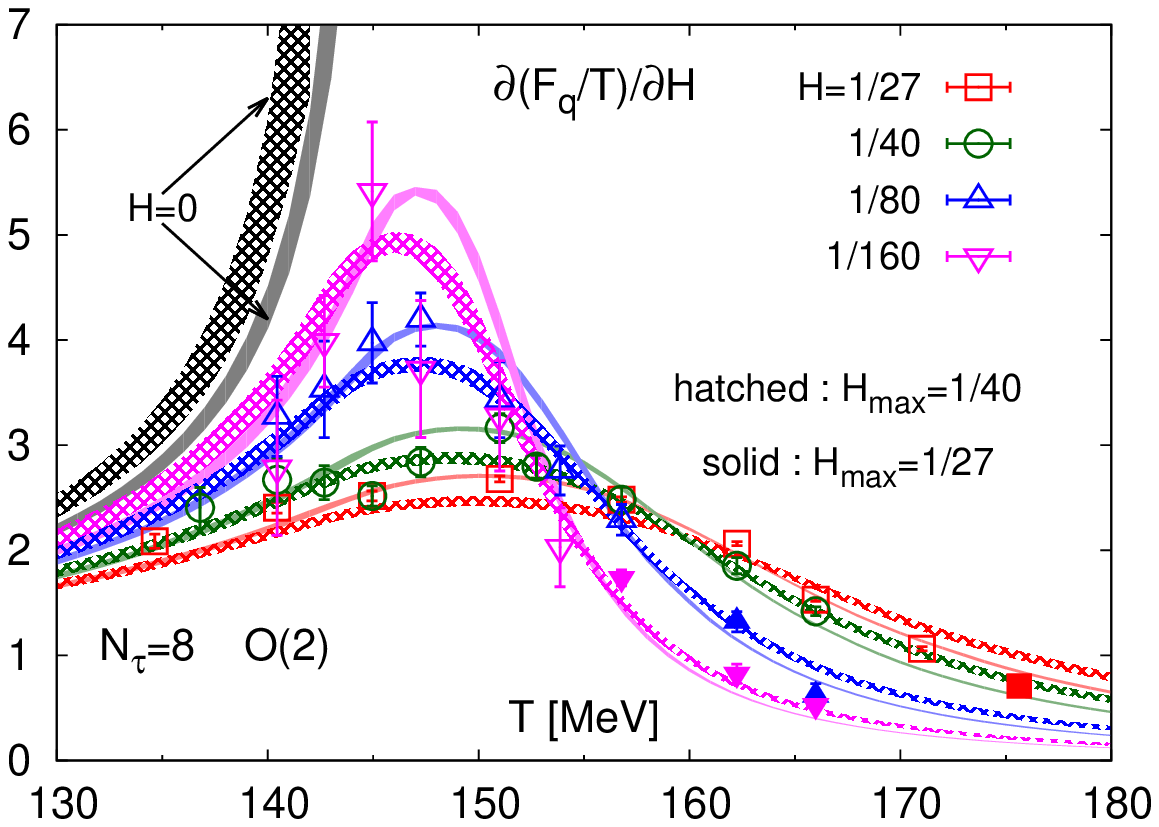}
\includegraphics[width=0.32\textwidth]{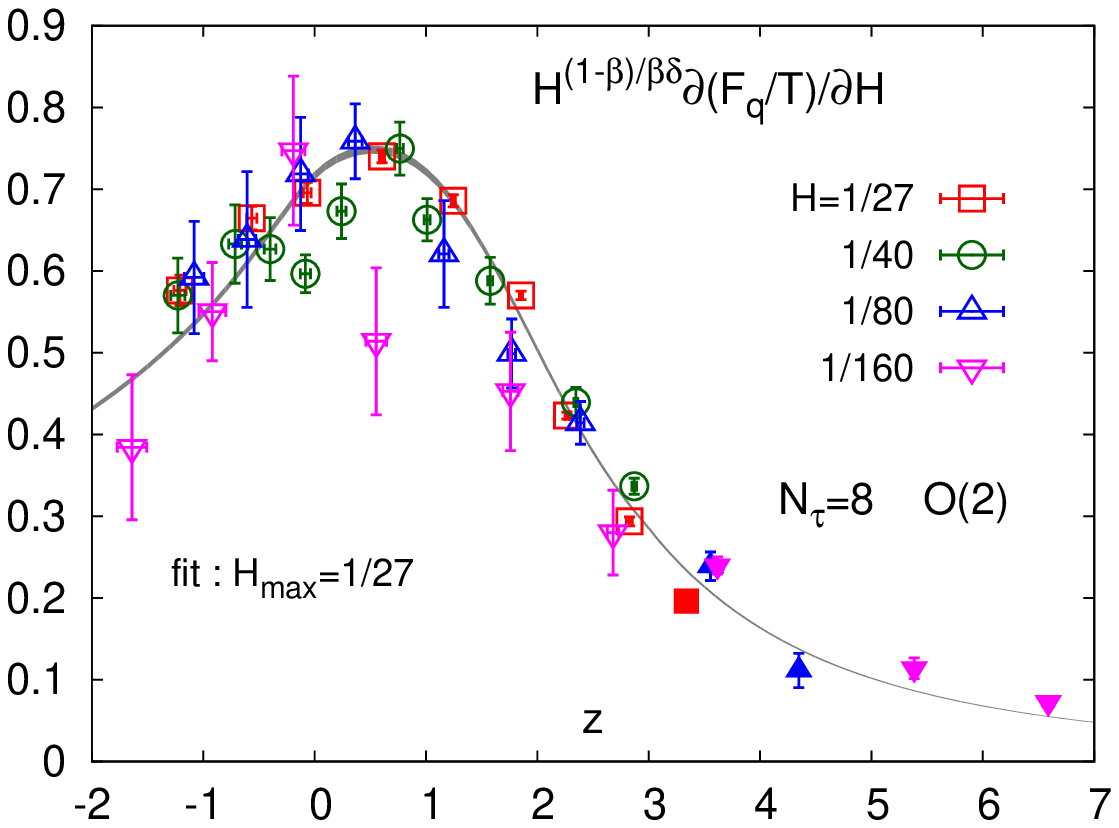}
\includegraphics[width=0.32\textwidth]{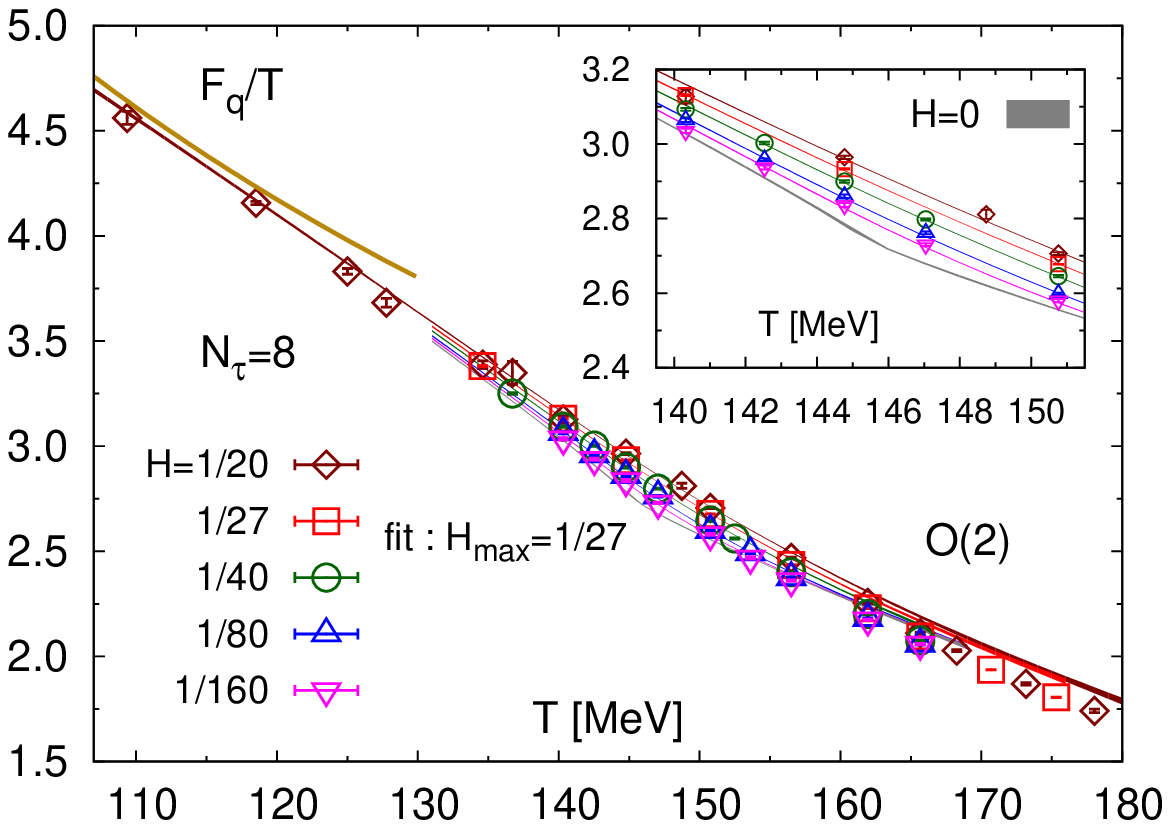}
\caption{Derivative of $F_q(T,H)/T$ with respect to $H$
  as function of $T$ (left) and the rescaled version for the fit with $\Hmax=1/27$ as function of $z$ (middle). Shown are results for several values
  of $m_l$, calculated on lattices with temporal extent $N_\tau=8$.   
  Curves in the left hand figure show results of 
  fits, with and without the data sets for $H=1/27$, based on the scaling ansatz given in eq.~\eqref{Fqm-crit},
  with fit parameters given in Table~\ref{tab:Fqo2fit}. Data shown with filled symbols correspond to $z\ge 3$ and have not been included in the fit. The right hand figure shows $F_q/T$ versus $T$. Fits are explained in the text.
  The inset in the right hand figure shows data in the temperature range
           covered by the fits. The chiral limit result for $H=0$ obtained
           from these fits is shown as grey bands (left, right). The solid gold line in the right hand figure
           shows the static-light meson contribution to $F_q/T$ 
           \cite{Bazavov:2013yv}.
  }
  \label{fig:FqdH}
\end{figure*}

Using the known asymptotic behavior of $f_f(z)$ for 
$z\rightarrow \pm \infty$ \cite{Engels:2011km} 
we obtain from eq.~\eqref{Fqcritical} for the quark mass 
dependence of  $F_q/T$  at fixed $T$ and for 
small $H$,
\begin{equation}
		\frac{F_q(T,H)}{T} \sim 
\begin{cases}
a^-(T)+A p_s^-(T)\ H &,\ T< T_c \\
a^r_{0,0}+A a_1\ H^{(1-\alpha)/\beta\delta} &,\ T=T_c \\
a^+(T)+p^+(T)\ H^2 &,\ T> T_c
\end{cases}
\; ,
\label{O4criticalmass}
\end{equation}
with $a^\pm(T)=A a_s^\pm(T)+\freg(T,0)$ as well as $p^+(T)=A p_s^+(T)+p^r_2(T)$ 
receiving contributions from both the singular and regular terms. 
For $T\le T_c$ the 
dominant quark mass dependence arises from the singular 
term only. In particular, we have
\begin{eqnarray}
\label{coefficients}
a_s^\pm(T) &=& (2-\alpha)\  z_0^{1-\alpha}\ c_0^{\pm}\ t |t|^{-\alpha} \; , \nonumber \\
p_s^-(T) &=& (2-\alpha-\beta\delta)\ (-z_0 t)^{1-\alpha- \beta\delta} \; ,  \\
p_s^+(T) &=& (2-\alpha- 2 \beta\delta)\ c_1^+ (z_0 t)^{1-\alpha-2 \beta\delta} \nonumber
\;,
\end{eqnarray}
where we followed the notation of Ref.~\cite{Engels:2011km} with $c_0^\pm$,
$c_1^+$
and $a_1$ denoting coefficients appearing in the parametrization of the 
scaling function $f_f(z)$.

For the  3-$d$, $\O(4)$ universality class these coefficients are given in 
\cite{Engels:2011km}. As we will present here only results from calculations for one non-zero value of the lattice spacing, $a=1/8T$, and do not perform a continuum extrapolation, we will use 
the critical exponents and scaling functions of the 3-$d$, $\O(2)$ 
universality class. We have extracted these from \cite{Engels:2000xw}, 
where the scaling functions are parametrized in the Widom-Griffiths form. 
We use $a_1=0.4734$, $c_0^-=2.447$,  $c_0^+ = 2.728$,  $c_1^+=-0.678$,
along with $\beta=0.349$, $\delta=4.780$ 
\cite{Hasenbusch:1999cc,Engels:2000xw,Clarke:2020clx}.

Making use of the relation between scaling functions, $f_f(z)$ 
and that of the order parameter, $f_G(z)$ (see for instance 
\cite{Engels:2011km}), 
we obtain in the vicinity of $T_c$,
\begin{equation}
\frac{\partial F_q(T,H)/T}{\partial H} = -A H^{(\beta -1)/\beta\delta}f'_G(z) 
             +\frac{\partial f_{\rm reg}(T,H)}{\partial H}\; .
\label{Fqm-crit}
\end{equation}
The derivative of $F_q/T$ with respect to $T$ is given by
\begin{equation}\label{eq:FqT}
T_c\frac{\partial F_q(T,H)/T}{\partial T} =  
A z_0 H^{-\alpha / \beta\delta} f''_f(z) + T_c \frac{\partial f_{\rm reg}(T,H)}{\partial T} \; .
\end{equation}
As $\alpha / \beta\delta < 0$ for the 3-$d$, $\O(2)$ as well
as $\O(4)$ universality classes, the derivatives of
$\ev{P}$ and $F_q/T$ with respect to $T$ do not diverge at $T_c$
in the limit $H\rightarrow 0$. 
On the other hand, the corresponding quark mass derivatives (eq.~\eqref{Fqm-crit})
are expected to diverge when approaching the chiral limit at $T_c$ 
as well as when approaching $T_c$ from below at $H=0$.
A possible influence of the chiral phase transition on the
quark mass dependence of $\ev{P}$ and $F_q/T$ is thus much 
easier to establish than on their temperature dependence. 


\emph{Computational setup and data analysis.---}
In this study we analyze properties of (2+1)-flavor QCD where 
the strange quark mass has been kept fixed to its physical value
and the two degenerate light quark masses are varied in the range 
$H=1/20$ to $1/160$, corresponding to a pion mass
$160~\text{MeV}\gtrsim m_\pi\gtrsim 58~\text{MeV}$ \cite{Ding:2019prx}.
The analysis is performed on sets of gauge field configurations that 
had been generated
previously by the HotQCD collaboration  
\cite{Bazavov:2011nk,Bazavov:2014pvz,Bazavov:2017dus,Ding:2019prx} on 
lattices of size $32^3\times 8$ ($H=1/20, 1/27$), $40^3\times 8$ 
($H=1/40$) and $56^3\times 8$ ($H=1/80,\ 1/160$).
using highly improved staggered quarks (HISQ) \cite{Follana:2006rc}
and the tree-level improved Symanzik gauge action.
These data have previously been
used in calculations of pseudo-critical temperatures \cite{Bazavov:2018mes} and the chiral phase transition temperature of (2+1)-flavor QCD \cite{Ding:2019prx}.
We also have generated additional configurations for $H=1/40,\ 1/80$. 
For the scale setting we use the kaon decay constant 
obtained in calculations with the HISQ action, {\it i.e.}
$f_K=156.1/\sqrt{2}$~MeV~\cite{Bazavov:2010hj}.

Observables have been calculated on lattices
with temporal extent $N_\tau=8$ and $N_\sigma/N_\tau = 4-7$.
For  $H =1/80$ we have results for the entire range of aspect ratios, $N_\sigma/N_\tau$. 
An analysis of finite volume effects on $\ev{P}$ shows no
significant volume dependence of  $\ev{P}$ in the entire
temperature range. Results for different $N_\sigma/N_\tau$
agree within errors
of about 1\%. We thus can safely neglect any finite volume
corrections to our results for $\ev{P}$. 
The statistics used in this study can be found in \cite{Clarke:2020clx},
along with more details for the finite volume dependence of $\ev{P}$.


\emph{Results.---}
In Fig.~\ref{fig:FqdH}~(left) we show results for the
derivative of $F_q(T,H)/T$ with respect to $H$ as function of $T$.
It is apparent  that  $\partial (F_q/T)/\partial H$
decreases with quark mass at high temperature and increases at low
temperature. This is consistent with an approach towards zero at high 
$T$ and a non-vanishing, strongly temperature dependent constant
at low $T$. Such a pattern is in accordance with the expected quadratic
dependence on $H$ for $T>T_c$ and the linear dependence for $T<T_c$
given in eq.~\eqref{O4criticalmass}.
Although errors are large for the results obtained with
the smallest light quark mass, $H=1/160$, it is evident that 
$\partial (F_q/T)/\partial H$ has maxima at
$T\sim 145-150$~MeV which are close to the chiral phase transition
temperature on lattices with temporal extent $N_\tau=8$ determined
in \cite{Ding:2019prx}, $T_c^{N_\tau=8}=144(2)$~MeV. With decreasing $H$ they
approach $T_c^{N_\tau=8}$ and the peak height increases.

In Fig.~\ref{fig:FqdH}~(middle) we show $\partial (F_q/T)/\partial H$
re\-scaled with the appropriate power of $H$ expected
from the $\O(2)$ scaling ansatz and plotted versus $z$. 
The good scaling behavior
suggests that $\partial (F_q/T)/\partial H$ will indeed diverge in the chiral limit and that $H$-dependent
contributions to $F_q/T$, arising from regular terms, are small compared to
those coming from the singular part. This motivated a fit ansatz for the
regular term that is independent of $H$, {\it i.e.} we use $\freg (T,H=0)$ in all our fits. 

A fit to $\partial (F_q/T)/\partial H$ thus only involves
the singular term of eq.~\eqref{Fqm-crit}. We performed this
3-parameter fit for all data sets with $H\le \Hmax$ by either including or leaving out the data for
$H=1/27$. These fits are shown in
Fig.~\ref{fig:FqdH}~(left), and the corresponding fit parameters,
$A$, $T_c$, $z_0$, are given in Table~\ref{tab:Fqo2fit}.
In particular, we find that $T_c$, obtained from these fits, as well as 
the non-universal scale parameter $z_0$, agree
well with earlier fit results for chiral susceptibilities
\cite{Ding:2019prx} in (2+1)-flavor QCD.
The rescaled fits are 
also shown in Fig.~\ref{fig:FqdH}~(middle).

\begin{figure}[t]
\centering
\includegraphics[width=0.32\textwidth]{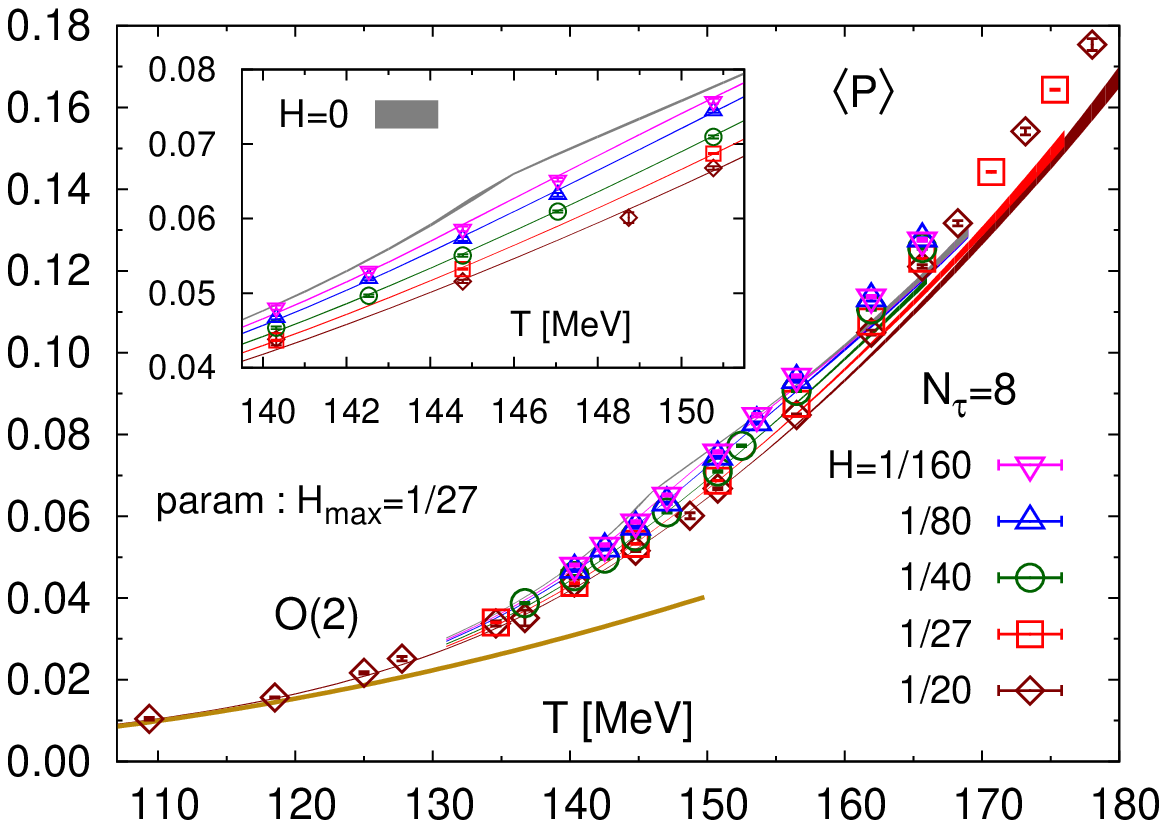}
\includegraphics[width=0.32\textwidth]{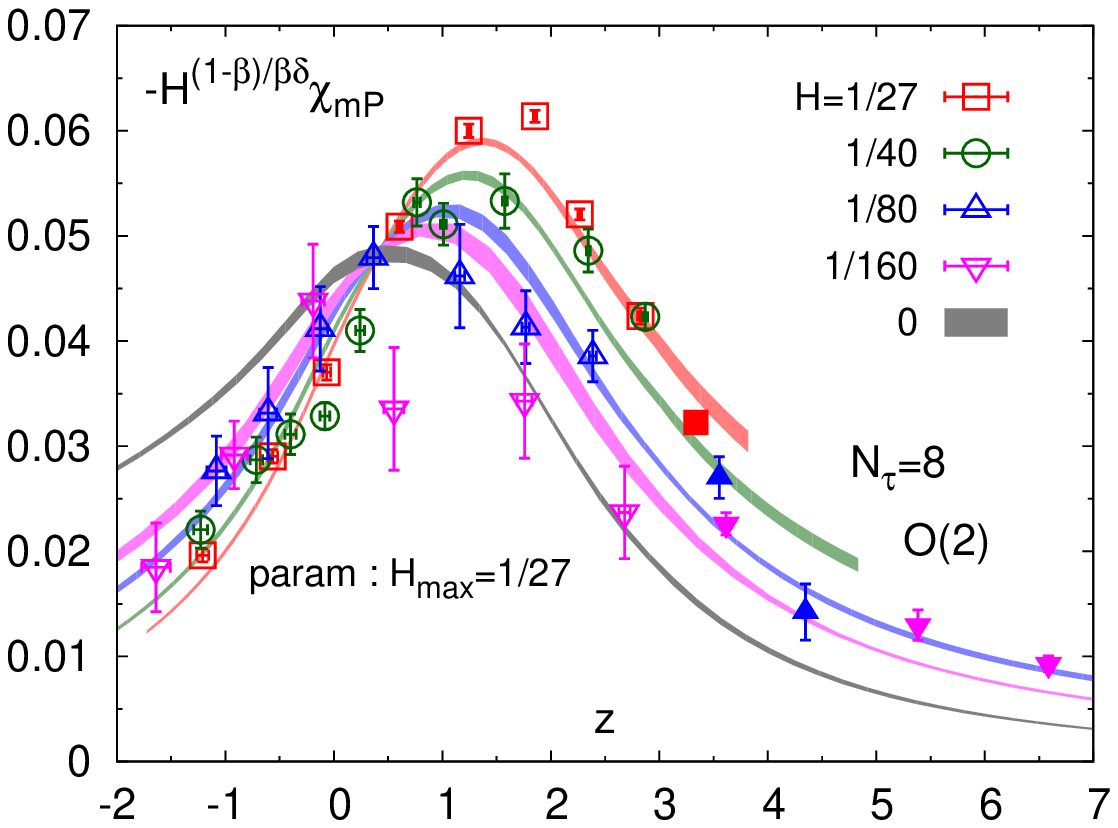}
  \caption{{\it Top:} $T$ dependence of $\ev{P}$
  obtained on lattices with temporal extent $N_\tau=8$
  for several values of $H$.
 The inset shows data in the $T$ range
           covered by the fit. Curves are based on a fit that used for the singular part parameters for $\Hmax=1/27$ given in Table~\ref{tab:Fqo2fit}. The chiral limit result for $H=0$ obtained
           from this fit is shown as a grey band. The solid gold 
	   line is as described in Fig.~\ref{fig:FqdH}.
           {\it Bottom:} Derivative of $\ev{P}$ versus scaling variable $z$.
           }
  \label{fig:PmassTall}
\end{figure}

\begin{table}
\centering
\begin{tabular}{cccc|cc||c} \hline\hline
 \multicolumn{4}{c|}{singular part}&\multicolumn{2}{c||}{regular part}&~\\
\hline
  $\Hmax$  &    $A$ & $T_c$ & $z_0$ & $a^r_{0,0}$ &  $a^r_{1,0}$ & $R^+$ \\
\hline
1/27 & 2.48(2)  & 145.6(3) & 2.24(5) & 2.74(1) & -34.4(7) & -0.92(1)\\
1/40 & 2.26(5)  & 144.2(6) & 1.83(9) & 2.81(3) & -27(1) &  -0.86(1)\\
\hline\hline
\end{tabular}
\caption{Summary of fit parameters for  $F_q(T,H)/T$
and the ratio $R^+$ introduced in eq.~\eqref{FqH02}.
The parameters for the singular part $(A,T_c,z_0)$ have
been obtained from a fit to $\partial (F_q/T)/\partial H$. 
}
        \label{tab:Fqo2fit}
\end{table}

Data for $F_q(T,H)/T$ are shown in Fig.~\ref{fig:FqdH} (right).
They have been fitted to the scaling
ansatz eq.~\eqref{Fqcritical} using only the 
constant and linear $H$-independent
terms in the regular part as fit parameters and keeping fixed
the three non-universal constants, determined in the 
previous step, in the singular part.
The $T$-range and data 
included in the fit are shown in the inset. The  resulting fit 
parameters, $a^r_{0,0}$, $a^r_{1,0}$,  are 
given in Table~\ref{tab:Fqo2fit}. In this figure we also show the 
static-light meson contribution to $F_q/T$ calculated in the hadron-gas approximation \cite{Megias:2012kb,Bazavov:2013yv}. In chiral perturbation theory this also 
gives a linear dependence on $H$ at low temperature \cite{Brambilla:2017hcq}.

\begin{figure}[t]
\centering
\includegraphics[width=0.32\textwidth]{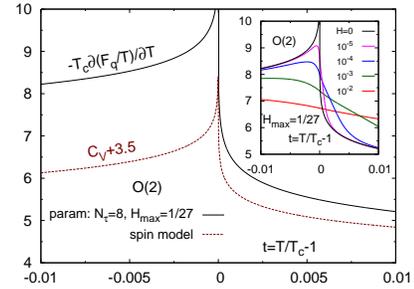}
\caption{Comparison of fits to the specific heat at $H=0$
of the 3-$d$, $\O(2)$
	     spin model taken from Ref.~\cite{Cucchieri:2002hu}
	     (dashed line) and the negative of the temperature
	     derivative of $F_q(T,0)/T$ (solid line). The 
	     former curve is shifted vertically by a constant 
	     for easier comparison with the spin model result.
         }
\label{fig:SpecificHeat}
\end{figure}

Once we have determined all five fit parameters for $F_q(T,H)/T$, we can plug them
into eq.~\eqref{Pcritical} to arrive at a parameter-free description of the $T$
and $H$ dependence of $\ev{P}$. The thus determined curves are shown in
Fig.~\ref{fig:PmassTall}. As seen in the inset, they agree well with $\ev{P}$
data near $T_c^{N_\tau=8}$, which suggests the behavior of $\ev{P}$ is explained
well by chiral scaling behavior in this region and serves
as a consistency check of our approach.

In Figs.~\ref{fig:FqdH}~(right) and \ref{fig:PmassTall}~(top) we also 
show the chiral limit ($H=0$) results for $F_q/T$ and $\ev{P}$. For the
former this is a sum of regular and singular contributions,
\begin{equation}
        \frac{F_q(T,0)}{T} = a^r_{0,0} + t \left( a^r_{1,0} + A^\pm |t|^{-\alpha} \right) \; ,
\label{FqH0}
\end{equation}
with $A^\pm=(2-\alpha) z_0^{1-\alpha} c_0^\pm A$ (see
eq.~\eqref{coefficients}).
Although at $T_c$ the contribution to the slope is entirely given by
the regular term $a_{1,0}^r$, close to $T_c$ this
contribution gets to a large extent canceled by
the singular contributions, $A^\pm |t|^{-\alpha}$. 
This is the origin of the well known spike in
specific-heat like observables ($2^{\rm nd}$
derivatives with respect to $T$) in the  $\O(N)$
universality classes. In the chiral limit our fit
results suggest the appearance of such a spike in the
temperature derivatives of  $F_q(T,0)/T$ as well as
$\ev{P}$. For the former we obtain
from eq.~\eqref{FqH0}, 
\begin{equation}
    T_c \frac{\partial (F_q(T,0)/T)}{\partial T} =  a^r_{1,0} \left( 1 + R^\pm  
       |t|^{-\alpha} \right)  \; ,
\label{FqH02}
\end{equation}
with $R^\pm=(1-\alpha)A^\pm/a^r_{1,0}$. $R^+$ is given in Table~\ref{tab:Fqo2fit} and 
$R^+/R^-= c^+_0/c^-_0=1.12(5)$ is a universal ratio \cite{Cucchieri:2002hu}. 
This makes it evident that already for $|t|=0.01$ the slope of $-F_q/T$ is about a factor 5 smaller than at $T_c$ . 

The basic features found in our analysis of 
(2+1)-flavor QCD are quite similar to those found in 
the analysis of 3-$d$, $\O(2)$ symmetric spin models 
\cite{Cucchieri:2002hu}. Also in that case a large 
cancellation of contributions arising from regular
and singular terms is found; the spike in the specific 
heat, $C_V$, is concentrated in a temperature interval 
of about 1\% around $T_c$, and $C_V$ changes by almost 
a factor $10$ in this temperature interval. In 
Fig.~\ref{fig:SpecificHeat} we show a comparison of the
scaling behavior of $-T_c\,\partial (F_q(T,0)/T)/\partial T$, 
obtained for the case of (2+1)-flavor QCD, and $C_V$ 
in $\O(2)$ spin models at $H=0$ \cite{Cucchieri:2002hu}.
The inset shows the development of this sharp peak as the quark mass 
decreases. It makes clear that this feature becomes visible only for
$H$ being substantially smaller than the region $H \simeq 10^{-2}$ that is
accessible in current lattice QCD calculations. 

\emph{Conclusions.--}
We have examined the quark mass dependence of the Polyakov loop 
expectation value and the heavy quark free energy extracted from it.
We provided evidence for the influence of 
chiral symmetry restoration that manifests in the singular behavior
of the quark mass derivatives of $\ev{P}$ and $F_q/T$ and arises from the 
energy-like 
behavior of these observables with respect to chiral transformations. 
These derivatives  diverge at $T_c$ in the chiral limit, consistent with
the expected behavior for energy-like observables in the 3-$d$, $\O(2)$ 
universality class. 

We showed that at finite values of the lattice spacing
the relative distribution between singular and regular contributions to 
the energy-like variables in QCD seems to be
similar to that in the 3-$d$, $\O(2)$ spin model. In particular, 
in the chiral limit the very narrow spike showing up in specific-heat like 
observables, which results
from a partial cancellation of singular and regular contributions combined
with the existence of a quite small, negative critical exponent $\alpha$,
is expected to show up also in the $T$-derivatives of $F_q/T$ and $\ev{P}$.

As the critical exponent $\alpha$ also is negative and small in the 
$\O(4)$ universality class, similar behavior of $F_q/T$ and $\ev{P}$
is expected to persist in the continuum limit.
 However, as $|\alpha|$ is an order of magnitude larger 
compared to the $\O(2)$ case, the spike may become broader
and may also be more prominent already for larger quark masses.
This may be tested also in effective model calculations.

All data from our calculations, presented in the figures of this paper, can
be found in \cite{Clarke:2020data}.

\section*{Acknowledgments}
This work was supported by the Deutsche Forschungsgemeinschaft
(DFG, German Research Foundation) Proj. No. 315477589-TRR 211; and by 
the German Bundesministerium f\"ur Bildung und Forschung through
Grant No. 05P18PBCA1.
Numerical calculations have been made possible through PRACE grants at CSCS, 
Switzerland, and grants at the Gauss Centre for Supercomputing and NIC-J\"ulich, Germany. These grants provided access to resources on Piz Daint at CSCS as well as on JUQUEEN and JUWELS at NIC. This research also used awards
of computer time at NERSC, Berkeley Laboratory, provided by the DOE ASCR 
Leadership Computing Challenge (ALCC) program. 
We thank the HotQCD Collaboration for providing access to their latest data
sets and for many fruitful discussions.

\bibliography{bibliography}

\begin{thebibliography}{27}%
\makeatletter
\providecommand \@ifxundefined [1]{%
 \@ifx{#1\undefined}
}%
\providecommand \@ifnum [1]{%
 \ifnum #1\expandafter \@firstoftwo
 \else \expandafter \@secondoftwo
 \fi
}%
\providecommand \@ifx [1]{%
 \ifx #1\expandafter \@firstoftwo
 \else \expandafter \@secondoftwo
 \fi
}%
\providecommand \natexlab [1]{#1}%
\providecommand \enquote  [1]{``#1''}%
\providecommand \bibnamefont  [1]{#1}%
\providecommand \bibfnamefont [1]{#1}%
\providecommand \citenamefont [1]{#1}%
\providecommand \href@noop [0]{\@secondoftwo}%
\providecommand \href [0]{\begingroup \@sanitize@url \@href}%
\providecommand \@href[1]{\@@startlink{#1}\@@href}%
\providecommand \@@href[1]{\endgroup#1\@@endlink}%
\providecommand \@sanitize@url [0]{\catcode `\\12\catcode `\$12\catcode
  `\&12\catcode `\#12\catcode `\^12\catcode `\_12\catcode `\%12\relax}%
\providecommand \@@startlink[1]{}%
\providecommand \@@endlink[0]{}%
\providecommand \url  [0]{\begingroup\@sanitize@url \@url }%
\providecommand \@url [1]{\endgroup\@href {#1}{\urlprefix }}%
\providecommand \urlprefix  [0]{URL }%
\providecommand \Eprint [0]{\href }%
\providecommand \doibase [0]{http://dx.doi.org/}%
\providecommand \selectlanguage [0]{\@gobble}%
\providecommand \bibinfo  [0]{\@secondoftwo}%
\providecommand \bibfield  [0]{\@secondoftwo}%
\providecommand \translation [1]{[#1]}%
\providecommand \BibitemOpen [0]{}%
\providecommand \bibitemStop [0]{}%
\providecommand \bibitemNoStop [0]{.\EOS\space}%
\providecommand \EOS [0]{\spacefactor3000\relax}%
\providecommand \BibitemShut  [1]{\csname bibitem#1\endcsname}%
\let\auto@bib@innerbib\@empty
\bibitem [{\citenamefont {Yaffe}\ and\ \citenamefont
  {Svetitsky}(1982)}]{Yaffe:1982qf}%
  \BibitemOpen
  \bibfield  {author} {\bibinfo {author} {\bibfnamefont {L.~G.}\ \bibnamefont
  {Yaffe}}\ and\ \bibinfo {author} {\bibfnamefont {B.}~\bibnamefont
  {Svetitsky}},\ }\href {\doibase 10.1103/PhysRevD.26.963} {\bibfield
  {journal} {\bibinfo  {journal} {Phys.\ Rev.\ D}\ }\textbf {\bibinfo {volume}
  {26}},\ \bibinfo {pages} {963} (\bibinfo {year} {1982})}\BibitemShut
  {NoStop}%
\bibitem [{\citenamefont {McLerran}\ and\ \citenamefont
  {Svetitsky}(1981{\natexlab{a}})}]{McLerran:1980pk}%
  \BibitemOpen
  \bibfield  {author} {\bibinfo {author} {\bibfnamefont {L.~D.}\ \bibnamefont
  {McLerran}}\ and\ \bibinfo {author} {\bibfnamefont {B.}~\bibnamefont
  {Svetitsky}},\ }\href {\doibase 10.1016/0370-2693(81)90986-2} {\bibfield
  {journal} {\bibinfo  {journal} {Phys.\ Lett.\ B}\ }\textbf {\bibinfo {volume}
  {98}},\ \bibinfo {pages} {195} (\bibinfo {year}
  {1981}{\natexlab{a}})}\BibitemShut {NoStop}%
\bibitem [{\citenamefont {Kuti}\ \emph {et~al.}(1981)\citenamefont {Kuti},
  \citenamefont {Polonyi},\ and\ \citenamefont {Szlachanyi}}]{Kuti:1980gh}%
  \BibitemOpen
  \bibfield  {author} {\bibinfo {author} {\bibfnamefont {J.}~\bibnamefont
  {Kuti}}, \bibinfo {author} {\bibfnamefont {J.}~\bibnamefont {Polonyi}}, \
  and\ \bibinfo {author} {\bibfnamefont {K.}~\bibnamefont {Szlachanyi}},\
  }\href {\doibase 10.1016/0370-2693(81)90987-4} {\bibfield  {journal}
  {\bibinfo  {journal} {Phys.\ Lett.\ B}\ }\textbf {\bibinfo {volume} {98}},\
  \bibinfo {pages} {199} (\bibinfo {year} {1981})}\BibitemShut {NoStop}%
\bibitem [{\citenamefont {Aoki}\ \emph {et~al.}(2009)\citenamefont {Aoki},
  \citenamefont {Borsanyi}, \citenamefont {Durr}, \citenamefont {Fodor},
  \citenamefont {Katz} \emph {et~al.}}]{Aoki:2009sc}%
  \BibitemOpen
  \bibfield  {author} {\bibinfo {author} {\bibfnamefont {Y.}~\bibnamefont
  {Aoki}}, \bibinfo {author} {\bibfnamefont {S.}~\bibnamefont {Borsanyi}},
  \bibinfo {author} {\bibfnamefont {S.}~\bibnamefont {Durr}}, \bibinfo {author}
  {\bibfnamefont {Z.}~\bibnamefont {Fodor}}, \bibinfo {author} {\bibfnamefont
  {S.~D.}\ \bibnamefont {Katz}},  \emph {et~al.},\ }\href {\doibase
  10.1088/1126-6708/2009/06/088} {\bibfield  {journal} {\bibinfo  {journal}
  {JHEP}\ }\textbf {\bibinfo {volume} {06}},\ \bibinfo {pages} {088} (\bibinfo
  {year} {2009})},\ \Eprint {http://arxiv.org/abs/0903.4155} {arXiv:0903.4155
  [hep-lat]} \BibitemShut {NoStop}%
\bibitem [{\citenamefont {Borsanyi}\ \emph {et~al.}(2010)\citenamefont
  {Borsanyi}, \citenamefont {Fodor}, \citenamefont {Hoelbling}, \citenamefont
  {Katz}, \citenamefont {Krieg} \emph {et~al.}}]{Borsanyi:2010bp}%
  \BibitemOpen
  \bibfield  {author} {\bibinfo {author} {\bibfnamefont {S.}~\bibnamefont
  {Borsanyi}}, \bibinfo {author} {\bibfnamefont {Z.}~\bibnamefont {Fodor}},
  \bibinfo {author} {\bibfnamefont {C.}~\bibnamefont {Hoelbling}}, \bibinfo
  {author} {\bibfnamefont {S.~D.}\ \bibnamefont {Katz}}, \bibinfo {author}
  {\bibfnamefont {S.}~\bibnamefont {Krieg}},  \emph {et~al.} (\bibinfo
  {collaboration} {Wuppertal-Budapest}),\ }\href {\doibase
  10.1007/JHEP09(2010)073} {\bibfield  {journal} {\bibinfo  {journal} {JHEP}\
  }\textbf {\bibinfo {volume} {09}},\ \bibinfo {pages} {073} (\bibinfo {year}
  {2010})},\ \Eprint {http://arxiv.org/abs/1005.3508} {arXiv:1005.3508
  [hep-lat]} \BibitemShut {NoStop}%
\bibitem [{\citenamefont {Bazavov}\ \emph {et~al.}(2016)\citenamefont
  {Bazavov}, \citenamefont {Brambilla}, \citenamefont {Ding}, \citenamefont
  {Petreczky}, \citenamefont {Schadler}, \citenamefont {Vairo},\ and\
  \citenamefont {Weber}}]{Bazavov:2016uvm}%
  \BibitemOpen
  \bibfield  {author} {\bibinfo {author} {\bibfnamefont {A.}~\bibnamefont
  {Bazavov}}, \bibinfo {author} {\bibfnamefont {N.}~\bibnamefont {Brambilla}},
  \bibinfo {author} {\bibfnamefont {H.~T.}\ \bibnamefont {Ding}}, \bibinfo
  {author} {\bibfnamefont {P.}~\bibnamefont {Petreczky}}, \bibinfo {author}
  {\bibfnamefont {H.~P.}\ \bibnamefont {Schadler}}, \bibinfo {author}
  {\bibfnamefont {A.}~\bibnamefont {Vairo}}, \ and\ \bibinfo {author}
  {\bibfnamefont {J.~H.}\ \bibnamefont {Weber}},\ }\href {\doibase
  10.1103/PhysRevD.93.114502} {\bibfield  {journal} {\bibinfo  {journal}
  {Phys.\ Rev.\ D}\ }\textbf {\bibinfo {volume} {93}},\ \bibinfo {pages}
  {114502} (\bibinfo {year} {2016})},\ \Eprint
  {http://arxiv.org/abs/1603.06637} {arXiv:1603.06637 [hep-lat]} \BibitemShut
  {NoStop}%
\bibitem [{\citenamefont {Clarke}\ \emph {et~al.}(2019)\citenamefont {Clarke},
  \citenamefont {Kaczmarek}, \citenamefont {Karsch},\ and\ \citenamefont
  {Lahiri}}]{Clarke:2019tzf}%
  \BibitemOpen
  \bibfield  {author} {\bibinfo {author} {\bibfnamefont {D.~A.}\ \bibnamefont
  {Clarke}}, \bibinfo {author} {\bibfnamefont {O.}~\bibnamefont {Kaczmarek}},
  \bibinfo {author} {\bibfnamefont {F.}~\bibnamefont {Karsch}}, \ and\ \bibinfo
  {author} {\bibfnamefont {A.}~\bibnamefont {Lahiri}},\ }\href@noop {}
  {\bibfield  {journal} {\bibinfo  {journal} {PoS}\ }\textbf {\bibinfo {volume}
  {LATTICE2019}},\ \bibinfo {pages} {194} (\bibinfo {year} {2019})},\ \Eprint
  {http://arxiv.org/abs/1911.07668} {arXiv:1911.07668 [hep-lat]} \BibitemShut
  {NoStop}%
\bibitem [{\citenamefont {Pisarski}\ and\ \citenamefont
  {Wilczek}(1984)}]{Pisarski:1983ms}%
  \BibitemOpen
  \bibfield  {author} {\bibinfo {author} {\bibfnamefont {R.~D.}\ \bibnamefont
  {Pisarski}}\ and\ \bibinfo {author} {\bibfnamefont {F.}~\bibnamefont
  {Wilczek}},\ }\href {\doibase 10.1103/PhysRevD.29.338} {\bibfield  {journal}
  {\bibinfo  {journal} {Phys.\ Rev.\ D}\ }\textbf {\bibinfo {volume} {29}},\
  \bibinfo {pages} {338} (\bibinfo {year} {1984})}\BibitemShut {NoStop}%
\bibitem [{\citenamefont {McLerran}\ and\ \citenamefont
  {Svetitsky}(1981{\natexlab{b}})}]{McLerran:1981pb}%
  \BibitemOpen
  \bibfield  {author} {\bibinfo {author} {\bibfnamefont {L.~D.}\ \bibnamefont
  {McLerran}}\ and\ \bibinfo {author} {\bibfnamefont {B.}~\bibnamefont
  {Svetitsky}},\ }\href {\doibase 10.1103/PhysRevD.24.450} {\bibfield
  {journal} {\bibinfo  {journal} {Phys.\ Rev.\ D}\ }\textbf {\bibinfo {volume}
  {24}},\ \bibinfo {pages} {450} (\bibinfo {year}
  {1981}{\natexlab{b}})}\BibitemShut {NoStop}%
\bibitem [{\citenamefont {Wilson}(1971{\natexlab{a}})}]{Wilson:1971bg}%
  \BibitemOpen
  \bibfield  {author} {\bibinfo {author} {\bibfnamefont {K.~G.}\ \bibnamefont
  {Wilson}},\ }\href {\doibase 10.1103/PhysRevB.4.3174} {\bibfield  {journal}
  {\bibinfo  {journal} {Phys. Rev. B}\ }\textbf {\bibinfo {volume} {4}},\
  \bibinfo {pages} {3174} (\bibinfo {year} {1971}{\natexlab{a}})}\BibitemShut
  {NoStop}%
\bibitem [{\citenamefont {Wilson}(1971{\natexlab{b}})}]{Wilson:1971dh}%
  \BibitemOpen
  \bibfield  {author} {\bibinfo {author} {\bibfnamefont {K.~G.}\ \bibnamefont
  {Wilson}},\ }\href {\doibase 10.1103/PhysRevB.4.3184} {\bibfield  {journal}
  {\bibinfo  {journal} {Phys. Rev. B}\ }\textbf {\bibinfo {volume} {4}},\
  \bibinfo {pages} {3184} (\bibinfo {year} {1971}{\natexlab{b}})}\BibitemShut
  {NoStop}%
\bibitem [{\citenamefont {Engels}\ and\ \citenamefont
  {Karsch}(2012)}]{Engels:2011km}%
  \BibitemOpen
  \bibfield  {author} {\bibinfo {author} {\bibfnamefont {J.}~\bibnamefont
  {Engels}}\ and\ \bibinfo {author} {\bibfnamefont {F.}~\bibnamefont
  {Karsch}},\ }\href {\doibase 10.1103/PhysRevD.85.094506} {\bibfield
  {journal} {\bibinfo  {journal} {Phys. Rev. D}\ }\textbf {\bibinfo {volume}
  {85}},\ \bibinfo {pages} {094506} (\bibinfo {year} {2012})},\ \Eprint
  {http://arxiv.org/abs/1105.0584} {arXiv:1105.0584 [hep-lat]} \BibitemShut
  {NoStop}%
\bibitem [{\citenamefont {Bazavov}\ and\ \citenamefont
  {Petreczky}(2013)}]{Bazavov:2013yv}%
  \BibitemOpen
  \bibfield  {author} {\bibinfo {author} {\bibfnamefont {A.}~\bibnamefont
  {Bazavov}}\ and\ \bibinfo {author} {\bibfnamefont {P.}~\bibnamefont
  {Petreczky}},\ }\href {\doibase 10.1103/PhysRevD.87.094505} {\bibfield
  {journal} {\bibinfo  {journal} {Phys.\ Rev.\ D}\ }\textbf {\bibinfo {volume}
  {87}},\ \bibinfo {pages} {094505} (\bibinfo {year} {2013})},\ \Eprint
  {http://arxiv.org/abs/1301.3943} {arXiv:1301.3943 [hep-lat]} \BibitemShut
  {NoStop}%
\bibitem [{\citenamefont {Engels}\ \emph {et~al.}(2000)\citenamefont {Engels},
  \citenamefont {Holtmann}, \citenamefont {Mendes},\ and\ \citenamefont
  {Schulze}}]{Engels:2000xw}%
  \BibitemOpen
  \bibfield  {author} {\bibinfo {author} {\bibfnamefont {J.}~\bibnamefont
  {Engels}}, \bibinfo {author} {\bibfnamefont {S.}~\bibnamefont {Holtmann}},
  \bibinfo {author} {\bibfnamefont {T.}~\bibnamefont {Mendes}}, \ and\ \bibinfo
  {author} {\bibfnamefont {T.}~\bibnamefont {Schulze}},\ }\href {\doibase
  10.1016/S0370-2693(00)01079-0} {\bibfield  {journal} {\bibinfo  {journal}
  {Phys. Lett. B}\ }\textbf {\bibinfo {volume} {492}},\ \bibinfo {pages} {219}
  (\bibinfo {year} {2000})},\ \Eprint {http://arxiv.org/abs/hep-lat/0006023}
  {arXiv:hep-lat/0006023} \BibitemShut {NoStop}%
\bibitem [{\citenamefont {Hasenbusch}\ and\ \citenamefont
  {Toeroek}(1999)}]{Hasenbusch:1999cc}%
  \BibitemOpen
  \bibfield  {author} {\bibinfo {author} {\bibfnamefont {M.}~\bibnamefont
  {Hasenbusch}}\ and\ \bibinfo {author} {\bibfnamefont {T.}~\bibnamefont
  {Toeroek}},\ }\href {\doibase 10.1088/0305-4470/32/36/301} {\bibfield
  {journal} {\bibinfo  {journal} {J. Phys. A}\ }\textbf {\bibinfo {volume}
  {32}},\ \bibinfo {pages} {6361} (\bibinfo {year} {1999})},\ \Eprint
  {http://arxiv.org/abs/cond-mat/9904408} {arXiv:cond-mat/9904408} \BibitemShut
  {NoStop}%
\bibitem [{\citenamefont {Clarke}\ \emph {et~al.}(2020)\citenamefont {Clarke},
  \citenamefont {Kaczmarek}, \citenamefont {Lahiri},\ and\ \citenamefont
  {Sarkar}}]{Clarke:2020clx}%
  \BibitemOpen
  \bibfield  {author} {\bibinfo {author} {\bibfnamefont {D.~A.}\ \bibnamefont
  {Clarke}}, \bibinfo {author} {\bibfnamefont {O.}~\bibnamefont {Kaczmarek}},
  \bibinfo {author} {\bibfnamefont {A.}~\bibnamefont {Lahiri}}, \ and\ \bibinfo
  {author} {\bibfnamefont {M.}~\bibnamefont {Sarkar}},\ }\href@noop {} {\
  (\bibinfo {year} {2020})},\ \Eprint {http://arxiv.org/abs/2010.15825}
  {arXiv:2010.15825 [hep-lat]} \BibitemShut {NoStop}%
\bibitem [{\citenamefont {Ding}\ \emph {et~al.}(2019)\citenamefont {Ding},
  \citenamefont {Hegde}, \citenamefont {Kaczmarek}, \citenamefont {Karsch},
  \citenamefont {Lahiri}, \citenamefont {Li} \emph {et~al.}}]{Ding:2019prx}%
  \BibitemOpen
  \bibfield  {author} {\bibinfo {author} {\bibfnamefont {H.~T.}\ \bibnamefont
  {Ding}}, \bibinfo {author} {\bibfnamefont {P.}~\bibnamefont {Hegde}},
  \bibinfo {author} {\bibfnamefont {O.}~\bibnamefont {Kaczmarek}}, \bibinfo
  {author} {\bibfnamefont {F.}~\bibnamefont {Karsch}}, \bibinfo {author}
  {\bibfnamefont {A.}~\bibnamefont {Lahiri}}, \bibinfo {author} {\bibfnamefont
  {S.-T.}\ \bibnamefont {Li}},  \emph {et~al.},\ }\href {\doibase
  10.1103/PhysRevLett.123.062002} {\bibfield  {journal} {\bibinfo  {journal}
  {Phys. Rev. Lett.}\ }\textbf {\bibinfo {volume} {123}},\ \bibinfo {pages}
  {062002} (\bibinfo {year} {2019})},\ \Eprint
  {http://arxiv.org/abs/1903.04801} {arXiv:1903.04801 [hep-lat]} \BibitemShut
  {NoStop}%
\bibitem [{\citenamefont {Bazavov}\ \emph {et~al.}(2012)\citenamefont
  {Bazavov}, \citenamefont {Bhattacharya}, \citenamefont {Cheng}, \citenamefont
  {DeTar}, \citenamefont {Ding}, \citenamefont {Gottlieb} \emph
  {et~al.}}]{Bazavov:2011nk}%
  \BibitemOpen
  \bibfield  {author} {\bibinfo {author} {\bibfnamefont {A.}~\bibnamefont
  {Bazavov}}, \bibinfo {author} {\bibfnamefont {T.}~\bibnamefont
  {Bhattacharya}}, \bibinfo {author} {\bibfnamefont {M.}~\bibnamefont {Cheng}},
  \bibinfo {author} {\bibfnamefont {C.}~\bibnamefont {DeTar}}, \bibinfo
  {author} {\bibfnamefont {H.~T.}\ \bibnamefont {Ding}}, \bibinfo {author}
  {\bibfnamefont {S.}~\bibnamefont {Gottlieb}},  \emph {et~al.},\ }\href
  {\doibase 10.1103/PhysRevD.85.054503} {\bibfield  {journal} {\bibinfo
  {journal} {Phys.\ Rev.\ D}\ }\textbf {\bibinfo {volume} {85}},\ \bibinfo
  {pages} {054503} (\bibinfo {year} {2012})},\ \Eprint
  {http://arxiv.org/abs/1111.1710} {arXiv:1111.1710 [hep-lat]} \BibitemShut
  {NoStop}%
\bibitem [{\citenamefont {Bazavov}\ \emph {et~al.}(2014)\citenamefont
  {Bazavov}, \citenamefont {Bhattacharya}, \citenamefont {DeTar}, \citenamefont
  {Ding}, \citenamefont {Gottlieb}, \citenamefont {Gupta} \emph
  {et~al.}}]{Bazavov:2014pvz}%
  \BibitemOpen
  \bibfield  {author} {\bibinfo {author} {\bibfnamefont {A.}~\bibnamefont
  {Bazavov}}, \bibinfo {author} {\bibfnamefont {T.}~\bibnamefont
  {Bhattacharya}}, \bibinfo {author} {\bibfnamefont {C.}~\bibnamefont {DeTar}},
  \bibinfo {author} {\bibfnamefont {H.~T.}\ \bibnamefont {Ding}}, \bibinfo
  {author} {\bibfnamefont {S.}~\bibnamefont {Gottlieb}}, \bibinfo {author}
  {\bibfnamefont {R.}~\bibnamefont {Gupta}},  \emph {et~al.} (\bibinfo
  {collaboration} {HotQCD}),\ }\href {\doibase 10.1103/PhysRevD.90.094503}
  {\bibfield  {journal} {\bibinfo  {journal} {Phys.\ Rev.\ D}\ }\textbf
  {\bibinfo {volume} {90}},\ \bibinfo {pages} {094503} (\bibinfo {year}
  {2014})},\ \Eprint {http://arxiv.org/abs/1407.6387} {arXiv:1407.6387
  [hep-lat]} \BibitemShut {NoStop}%
\bibitem [{\citenamefont {Bazavov}\ \emph {et~al.}(2017)\citenamefont
  {Bazavov}, \citenamefont {Ding}, \citenamefont {Hegde}, \citenamefont
  {Kaczmarek}, \citenamefont {Karsch}, \citenamefont {Laermann} \emph
  {et~al.}}]{Bazavov:2017dus}%
  \BibitemOpen
  \bibfield  {author} {\bibinfo {author} {\bibfnamefont {A.}~\bibnamefont
  {Bazavov}}, \bibinfo {author} {\bibfnamefont {H.~T.}\ \bibnamefont {Ding}},
  \bibinfo {author} {\bibfnamefont {P.}~\bibnamefont {Hegde}}, \bibinfo
  {author} {\bibfnamefont {O.}~\bibnamefont {Kaczmarek}}, \bibinfo {author}
  {\bibfnamefont {F.}~\bibnamefont {Karsch}}, \bibinfo {author} {\bibfnamefont
  {E.}~\bibnamefont {Laermann}},  \emph {et~al.},\ }\href {\doibase
  10.1103/PhysRevD.95.054504} {\bibfield  {journal} {\bibinfo  {journal} {Phys.
  Rev. D}\ }\textbf {\bibinfo {volume} {95}},\ \bibinfo {pages} {054504}
  (\bibinfo {year} {2017})},\ \Eprint {http://arxiv.org/abs/1701.04325}
  {arXiv:1701.04325 [hep-lat]} \BibitemShut {NoStop}%
\bibitem [{\citenamefont {Follana}\ \emph {et~al.}(2007)\citenamefont
  {Follana}, \citenamefont {Mason}, \citenamefont {Davies}, \citenamefont
  {Hornbostel}, \citenamefont {Lepage} \emph {et~al.}}]{Follana:2006rc}%
  \BibitemOpen
  \bibfield  {author} {\bibinfo {author} {\bibfnamefont {E.}~\bibnamefont
  {Follana}}, \bibinfo {author} {\bibfnamefont {Q.}~\bibnamefont {Mason}},
  \bibinfo {author} {\bibfnamefont {C.}~\bibnamefont {Davies}}, \bibinfo
  {author} {\bibfnamefont {K.}~\bibnamefont {Hornbostel}}, \bibinfo {author}
  {\bibfnamefont {P.}~\bibnamefont {Lepage}},  \emph {et~al.} (\bibinfo
  {collaboration} {HPQCD, UKQCD}),\ }\href {\doibase
  10.1103/PhysRevD.75.054502} {\bibfield  {journal} {\bibinfo  {journal} {Phys.
  Rev. D}\ }\textbf {\bibinfo {volume} {75}},\ \bibinfo {pages} {054502}
  (\bibinfo {year} {2007})},\ \Eprint {http://arxiv.org/abs/hep-lat/0610092}
  {arXiv:hep-lat/0610092} \BibitemShut {NoStop}%
\bibitem [{\citenamefont {Bazavov}\ \emph {et~al.}(2019)\citenamefont
  {Bazavov}, \citenamefont {Ding}, \citenamefont {Hegde}, \citenamefont
  {Kaczmarek}, \citenamefont {Karsch}, \citenamefont {Karthik} \emph
  {et~al.}}]{Bazavov:2018mes}%
  \BibitemOpen
  \bibfield  {author} {\bibinfo {author} {\bibfnamefont {A.}~\bibnamefont
  {Bazavov}}, \bibinfo {author} {\bibfnamefont {H.~T.}\ \bibnamefont {Ding}},
  \bibinfo {author} {\bibfnamefont {P.}~\bibnamefont {Hegde}}, \bibinfo
  {author} {\bibfnamefont {O.}~\bibnamefont {Kaczmarek}}, \bibinfo {author}
  {\bibfnamefont {F.}~\bibnamefont {Karsch}}, \bibinfo {author} {\bibfnamefont
  {N.}~\bibnamefont {Karthik}},  \emph {et~al.} (\bibinfo {collaboration}
  {HotQCD}),\ }\href {\doibase 10.1016/j.physletb.2019.05.013} {\bibfield
  {journal} {\bibinfo  {journal} {Phys.\ Lett.\ B}\ }\textbf {\bibinfo {volume}
  {795}},\ \bibinfo {pages} {15} (\bibinfo {year} {2019})},\ \Eprint
  {http://arxiv.org/abs/1812.08235} {arXiv:1812.08235 [hep-lat]} \BibitemShut
  {NoStop}%
\bibitem [{\citenamefont {Bazavov}\ \emph {et~al.}(2010)\citenamefont
  {Bazavov}, \citenamefont {Bernard}, \citenamefont {DeTar}, \citenamefont
  {Du}, \citenamefont {Freeman}, \citenamefont {Gottlieb} \emph
  {et~al.}}]{Bazavov:2010hj}%
  \BibitemOpen
  \bibfield  {author} {\bibinfo {author} {\bibfnamefont {A.}~\bibnamefont
  {Bazavov}}, \bibinfo {author} {\bibfnamefont {C.}~\bibnamefont {Bernard}},
  \bibinfo {author} {\bibfnamefont {C.}~\bibnamefont {DeTar}}, \bibinfo
  {author} {\bibfnamefont {X.}~\bibnamefont {Du}}, \bibinfo {author}
  {\bibfnamefont {W.}~\bibnamefont {Freeman}}, \bibinfo {author} {\bibfnamefont
  {S.}~\bibnamefont {Gottlieb}},  \emph {et~al.} (\bibinfo {collaboration}
  {MILC}),\ }\href {\doibase 10.22323/1.105.0074} {\bibfield  {journal}
  {\bibinfo  {journal} {PoS}\ }\textbf {\bibinfo {volume} {LATTICE2010}},\
  \bibinfo {pages} {074} (\bibinfo {year} {2010})},\ \Eprint
  {http://arxiv.org/abs/1012.0868} {arXiv:1012.0868 [hep-lat]} \BibitemShut
  {NoStop}%
\bibitem [{\citenamefont {Megias}\ \emph {et~al.}(2012)\citenamefont {Megias},
  \citenamefont {Arriola},\ and\ \citenamefont {Salcedo}}]{Megias:2012kb}%
  \BibitemOpen
  \bibfield  {author} {\bibinfo {author} {\bibfnamefont {E.}~\bibnamefont
  {Megias}}, \bibinfo {author} {\bibfnamefont {E.~R.}\ \bibnamefont {Arriola}},
  \ and\ \bibinfo {author} {\bibfnamefont {L.~L.}\ \bibnamefont {Salcedo}},\
  }\href {\doibase 10.1103/PhysRevLett.109.151601} {\bibfield  {journal}
  {\bibinfo  {journal} {Phys.\ Rev.\ Lett.}\ }\textbf {\bibinfo {volume}
  {109}},\ \bibinfo {pages} {151601} (\bibinfo {year} {2012})},\ \Eprint
  {http://arxiv.org/abs/1204.2424} {arXiv:1204.2424 [hep-ph]} \BibitemShut
  {NoStop}%
\bibitem [{\citenamefont {Brambilla}\ \emph {et~al.}(2018)\citenamefont
  {Brambilla}, \citenamefont {Komijani}, \citenamefont {Kronfeld},\ and\
  \citenamefont {Vairo}}]{Brambilla:2017hcq}%
  \BibitemOpen
  \bibfield  {author} {\bibinfo {author} {\bibfnamefont {N.}~\bibnamefont
  {Brambilla}}, \bibinfo {author} {\bibfnamefont {J.}~\bibnamefont {Komijani}},
  \bibinfo {author} {\bibfnamefont {A.~S.}\ \bibnamefont {Kronfeld}}, \ and\
  \bibinfo {author} {\bibfnamefont {A.}~\bibnamefont {Vairo}} (\bibinfo
  {collaboration} {TUMQCD}),\ }\href {\doibase 10.1103/PhysRevD.97.034503}
  {\bibfield  {journal} {\bibinfo  {journal} {Phys. Rev. D}\ }\textbf {\bibinfo
  {volume} {97}},\ \bibinfo {pages} {034503} (\bibinfo {year} {2018})},\
  \Eprint {http://arxiv.org/abs/1712.04983} {arXiv:1712.04983 [hep-ph]}
  \BibitemShut {NoStop}%
\bibitem [{\citenamefont {Cucchieri}\ \emph {et~al.}(2002)\citenamefont
  {Cucchieri}, \citenamefont {Engels}, \citenamefont {Holtmann}, \citenamefont
  {Mendes},\ and\ \citenamefont {Schulze}}]{Cucchieri:2002hu}%
  \BibitemOpen
  \bibfield  {author} {\bibinfo {author} {\bibfnamefont {A.}~\bibnamefont
  {Cucchieri}}, \bibinfo {author} {\bibfnamefont {J.}~\bibnamefont {Engels}},
  \bibinfo {author} {\bibfnamefont {S.}~\bibnamefont {Holtmann}}, \bibinfo
  {author} {\bibfnamefont {T.}~\bibnamefont {Mendes}}, \ and\ \bibinfo {author}
  {\bibfnamefont {T.}~\bibnamefont {Schulze}},\ }\href {\doibase
  10.1088/0305-4470/35/31/301} {\bibfield  {journal} {\bibinfo  {journal} {J.
  Phys. A}\ }\textbf {\bibinfo {volume} {35}},\ \bibinfo {pages} {6517}
  (\bibinfo {year} {2002})},\ \Eprint {http://arxiv.org/abs/cond-mat/0202017}
  {arXiv:cond-mat/0202017} \BibitemShut {NoStop}%
\bibitem [{\citenamefont {Clarke}\ \emph {et~al.}(2021)\citenamefont {Clarke},
  \citenamefont {Kaczmarek}, \citenamefont {Lahiri},\ and\ \citenamefont
  {Sarkar}}]{Clarke:2020data}%
  \BibitemOpen
  \bibfield  {author} {\bibinfo {author} {\bibfnamefont {D.~A.}\ \bibnamefont
  {Clarke}}, \bibinfo {author} {\bibfnamefont {O.}~\bibnamefont {Kaczmarek}},
  \bibinfo {author} {\bibfnamefont {A.}~\bibnamefont {Lahiri}}, \ and\ \bibinfo
  {author} {\bibfnamefont {M.}~\bibnamefont {Sarkar}},\ }\href {\doibase
  10.4119/unibi/2950112} {\  (\bibinfo {year} {2021}),\
  10.4119/unibi/2950112}\BibitemShut {NoStop}%
\end{thebibliography}%

\end{document}